# Observation of antiferromagnetic order collapse in the pressurized insulator LaMnPO

Jing Guo[1], J. W. Simonson[2], Liling Sun[1]*, Qi Wu[1], Peiwen Gao[1], Chao Zhang[1], Dachun Gu[1], Gabriel Kotliar[3], Meigan Aronson[2] and Zhongxian Zhao[1]*

[1]*Institute of Physics and Beijing National Laboratory for Condensed Matter Physics, Chinese Academy of Sciences, Beijing 100190, China*
[2]*Department of Physics and Astronomy, Stony Brook University, Stony Brook, NY 11794, USA*
[3]*Department of Physics and Astronomy, Rutgers University, Piscataway, NJ 08854, USA*

The emergence of superconductivity in the iron pnictide and cuprate high temperature superconductors usually accompanies the suppression of an antiferromagnetically (AFM) ordered state in a corresponding parent compound through the use of chemical doping or external pressure[1-5]. A great deal of effort has been made to find superconductivity in Mn-based compounds[6-14], which are thought to bridge the gap between the two families of high temperature superconductors[7,15,16], but long-ranged AFM order was not successfully suppressed via chemical doping in these investigations. Here we report the first observation of the pressure-induced elimination of long-ranged AFM order in LaMnPO single crystals that are iso-structural to the LaFeAsO superconductor[15,17]. By combining *in-situ* high pressure resistance and *ac* susceptibility measurements, we found that LaMnPO undergoes a crossover from an AFM insulating to an AFM metallic state at a pressure ~20 GPa and that the long-ranged AFM order collapses at a higher pressure ~32 GPa. Our findings are of importance to explore potential superconductivity in Mn-based **compounds and to shed new light on the underlying mechanism of high temperature superconductivity.**

As is found in other Mn-based pnictide compounds that are iso-structural to the parent compound of iron pnictide superconductor LaFeAsO[15,17], the physical properties of LaMnPO bear a strong resemblance to those of the parent compounds of cuprate high temperature superconductors[2]. At ambient pressure, LaMnPO is a localized moment AFM insulator with a Néel temperature ($T_N$) about 375 K, and the ~3.2 $\mu_B$ per Mn moments align antiferromagnetically in a checkerboard pattern that is stacked along the $c$ axis [15,16]. Therefore, it is appropriate to study LaMnPO with the aim of uncovering the underlying mechanism of high temperature superconductivity and for seeking potential superconductivity. Like all doping studies of related $BaMn_2As_2$[18-21] and $CaMn_2Sb_2$[13], long-ranged AFM order in LaMnPO is also robust upon chemical doping. The substitution of fluorine for oxygen to its solubility limit does not much alter the long-ranged AFM order state, and no metallization is observed in $LaMnP(O_{1-x}F_x)$[14,16]. All results that are available so far demonstrate that chemical doping does not have a decisive influence on the long-ranged AFM order in these Mn-based compounds. High-pressure studies on $BaMn_2As_2$ find evidence of metallization[22], however, no experimental investigation of the variation of the long-range AFM order with pressure has been reported. In this study, we applied high pressure techniques to probe the possible elimination of the long-range AFM order and the potential for superconductivity in LaMnPO.

Figure 1(a) displays the temperature ($T$) dependence of the electrical resistance ($R$)

of a LaMnPO single crystal under different pressures. Instrumental restrictions limit our measurement of full *R-T* curves to pressures larger than 11.7 GPa. Here, the low temperature resistances show a huge upturn, which is suppressed remarkably with further increasing of the pressure. Significantly, the insulating behavior at lower temperatures is completely suppressed and metallic behavior emerges (as indicated by arrows) at pressures of 20.8 GPa and above, indicating the occurrence of partial electronic delocalization in LaMnPO. To determine the critical pressure for the crossover from the insulating to the metallic states in LaMnPO precisely, we plot the pressure dependencies of the resistance measured at different temperatures (Fig.1 b). It is noted that the temperature dependence of the resistance changes its trend at a pressure between 19.6 and 20.8 GPa. Thus, we take the average value (20.3 GPa) as the critical pressure $P_C$ for the T=0 insulator-metal transition. Careful inspection of the *R-T* curves measured at pressures between 20.8 GPa and 25.5 GPa, finds a sizable maximum centered at a temperature T', revealing that there exists an intermediary phase around the crossover pressure $P_C$ in compressed LaMnPO (Fig.1c). The hump signals that the itinerant and localized electrons coexist and compete with each other over this intermediate pressure and temperature regime. The hump disappears when the pressure is increased to 25.5 GPa. At pressures above 31.2 GPa, the resistance becomes linearly proportional to temperature (Fig.1c), indicating that the system enters a pure metallic state.

Figure 2a shows the resistance as a function of the reciprocal temperature for the LaMnPO single crystal subjected to different pressures. We found that these Arrhenius plots are linear at high temperatures for pressures as large as 25.5 GPa. On basis of the Arrhenius equation, $\rho \sim exp(\varepsilon_A/2k_BT)$, we can plot the activation energy for the excitation of charge carriers ($\varepsilon_A$) as a function of pressure for the sample investigated. As shown in Fig.2b, $\varepsilon_A$ decreases rapidly with increasing pressure below 15 GPa, where the insulating behavior is systematically suppressed (Fig.1a). It is noted that $\varepsilon_A$ remains unchanged with a value of about 60 meV in the pressure range 16-20 GPa. We attribute this gap to the competition between the itinerant and localized electrons. As pressure is elevated above 20.3 GPa, the $\varepsilon_A$ continues to decline and the metallicity of the system emerges (inset of Fig.2b). At pressure above 32 GPa, we found that $\varepsilon_A$ approaches zero, demonstrating that the gap has collapsed.

It is known from studies of cuprate and iron pnictide superconductors that the full suppression of long-ranged AFM order is a crucial step to realize superconductivity, and one can ask whether this may also be the case for LaMnPO. We probed the evolution of AFM order with pressure by measuring the onset temperature $T_N$ by means of *in-situ* high pressure *ac* susceptibility. Since the ambient pressure value of $T_N$ of LaMnPO is about 375 K[15,16], we cannot detect $T_N$ at pressures less than 7.3 GPa, where the AFM transition becomes apparent as shown in Fig.3. With increasing pressure, we found that $T_N$ shifts to lower temperatures and then vanishes at ~32 GPa, indicating that pressure destroys the long-range AFM order and causes the system to undergo an AFM-PM transition where all electrons related to the AFM order are delocalized. This is in excellent agreement with our high-pressure resistance data. At pressures above 32 GPa, we can see that the system exhibits metallic behavior over the entire temperature range (Fig. 1c). This is the first observation of the collapse of long-ranged AFM order in a Mn-based compound that is iso-structural to the iron pnictide superconductors, an event that was predicted by our previous calculations[16].

The overall behavior of pressurized LaMnPO is summarized in the electronic phase diagram presented in Fig.4. $T_N$ is indicated by filled triangles, and we find it to be systematically suppressed upon increasing pressure. At ~19 GPa, the system enters a mixed state (referred to as the M' phase), in which the insulating (I) and metallic (M) states coexist (Fig.1c), while long-ranged AFM order is still retained. The AFM-M' regime that is delineated by the resistance hump lies in the pressure range 19-32 GPa and develops from a high temperature state that is definitively insulating. At pressures above 32 GPa, there is no experimental indication of AFM order, demonstrating that the long-ranged AFM order has collapsed and that the localized/moment-bearing electrons are now fully delocalized. We note that the system only becomes fully metallic at pressures where the long-ranged AFM order at T=0 has collapsed (Fig.1c), implying that the insulating behavior of the intermediary phase requires at least some form of AFM short range order. These results, and in particular the resistance hump observed at intermediate pressures, are reminiscent of the experimental findings in compressed $A_2Fe_4Se_5$ (A=K or Tl substituted on Rb )[23]. Coexistence of localized and itinerant carriers has been suggested for these systems [23-26]. In the present study, no superconductivity is found in LaMnPO above the temperature of 1.5 K. However, the observations of the long-ranged AFM order collapse and the novel intermediary phase (M' phase) provide fresh information for exploring superconductivity in Mn-based compounds and provide new insight into the underlying mechanism of high temperature superconductivity.

**Method**

The single crystals of LaMnPO were grown by a NaCl-KCl eutectic flux method, as described in Ref 14 and 27. Diamond anvil cells (DACs) were used to create high pressure. The anvil diameter is about 300 µm. High-pressure electrical resistance experiments were carried out in a DAC using a standard four-probe technique. High-pressure alternating current (*ac*) susceptibility measurements were conducted using home-made coils that were wound around a diamond anvil[28,29]. The nonmagnetic rhenium gasket was preindented down to a 50µm thickness for different runs of high-pressure resistance and magnetic measurements. A thin plate cleaved from the LaMnPO single crystal with dimensions of around 80×80×10 µm was loaded into the gasket hole in a DAC for the resistance measurements. To detect the variation of $T_N$ with pressure more clearly, several thin plates with the same dimensions were loaded into another gasket hole sitting in a different DAC for *ac* susceptibility measurements. No pressure medium was used in the high-pressure magnetic or resistance measurements. Pressure was determined by ruby fluorescence[30].

**Acknowledgements**

Work in China was supported by the NSCF (Grant No. 11074294), 973 projects (Grant No. 2011CBA00100 and 2010CB923000) and Chinese Academy of Sciences. Research at Stony Brook (J. W. S and M. C. A) and Rutgers (G. K) was supported by a Department of Defense National Security Science and Engineering Faculty Fellowship via the Air Force Office of Scientific Research.

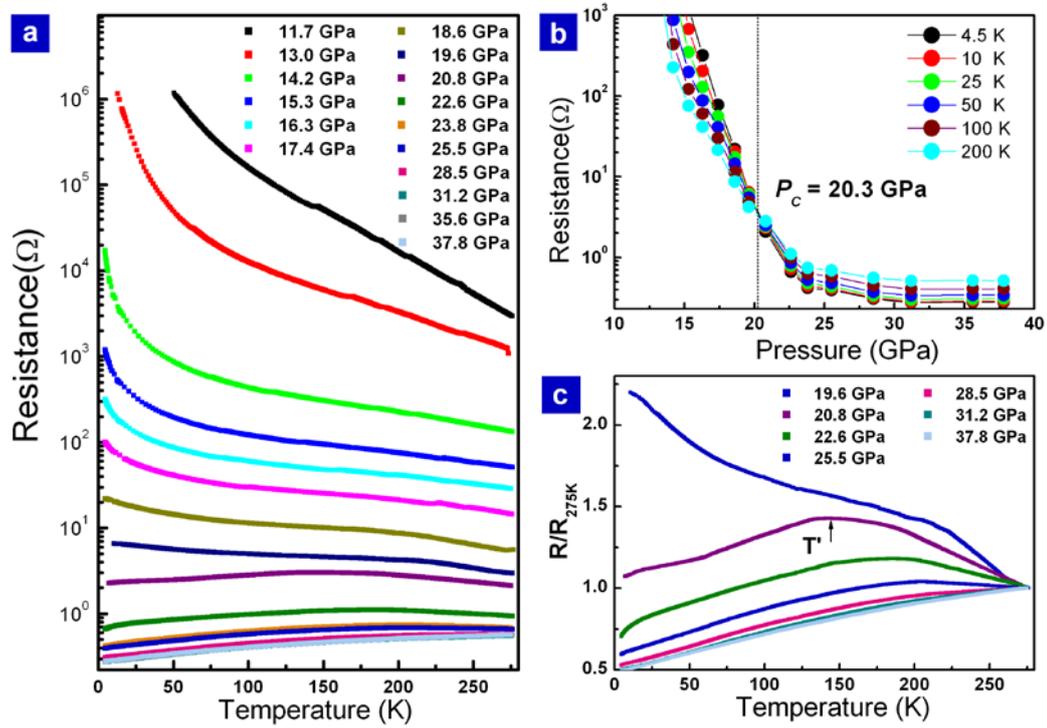

**Figure 1 Temperature dependencies of the electrical resistance of a single crystal of LaMnPO, measured at different pressures, and as a function of pressure at different fixed temperatures.** (a) Resistance-Temperature (*R-T*) curves measured in the pressure range 11.7-37.8 GPa, display a remarkable pressure-induced suppression of the insulating behavior. (b) The pressure dependence of the resistance obtained at different fixed temperatures, exhibiting a critical pressure ($P_C$) of the transition from the insulating state to metallic state in LaMnPO. (c) Selected *R-T* curves measured in the pressure range 19.6-37.8 GPa, show an intermediary phase featuring a resistance hump that is driven by pressure. The temperature scale T′ represents the temperature with the maximum resistance, as indicated by the arrow. This hump phenomenon lies in the pressure range 19.6- 25.5 GPa, and is completely suppressed at 28.5 GPa and above, where the system enters into a fully metallic state.

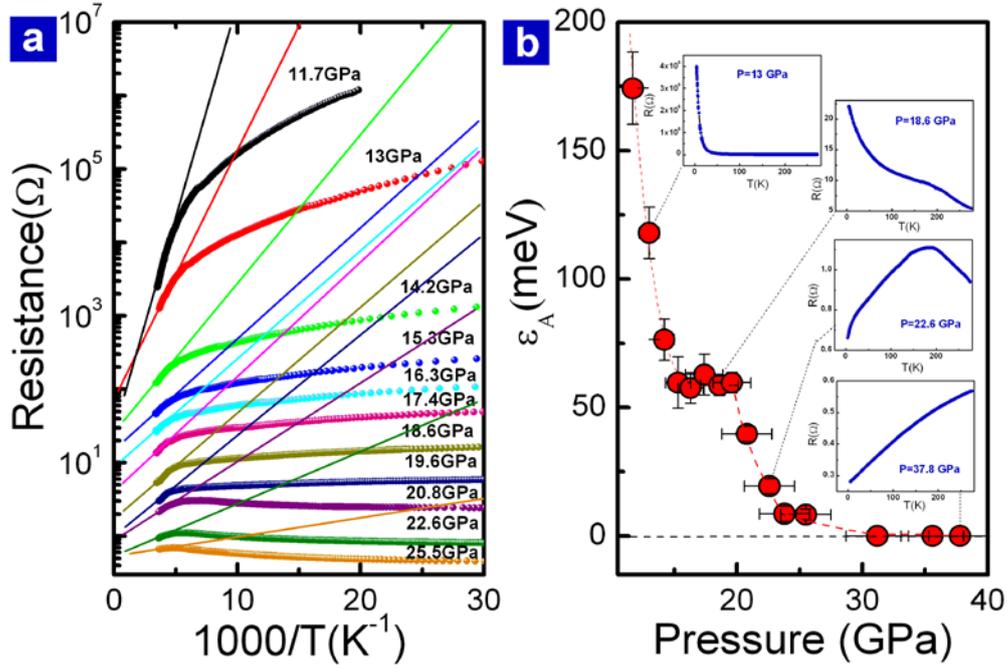

**Figure 2 Arrhenius plots of the temperature dependence of the resistance at different pressures for a LaMnPO single crystal and its activation energy gap ε$_A$ as a function of pressure.** (a) Resistance as a function of the reciprocal temperature, measured at different pressures. Solid lines are the best high-temperature fits. (b) The pressure dependence of the activation energy gap ε$_A$, which shrinks rapidly with increasing pressure below 15 GPa, remains almost unchanged in the pressure range 15-20 GPa, declines with further increases of pressure, and finally vanishes at a pressure of 32 GPa. The insets display the corresponding *R-T* curves measured at representative pressures.

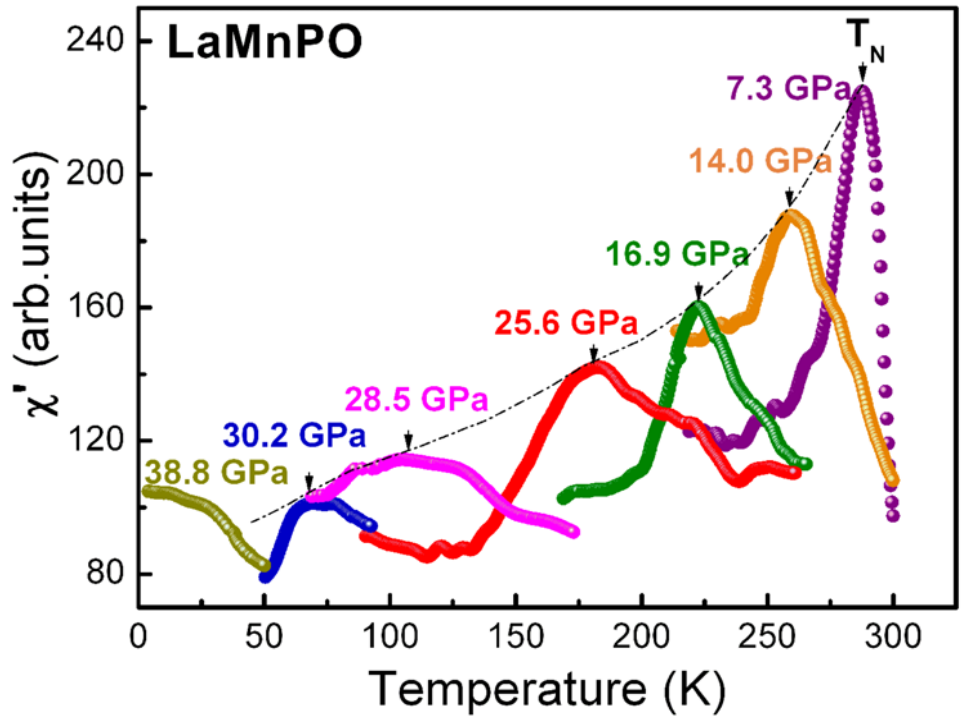

**Figure 3 Temperature dependencies of the real part of the alternating current (*ac*) susceptibility χ' measured at different pressures for the LaMnPO single crystal.** The arrows indicate maxima in χ'(T) that mark the onset of AFM order at the Néel temperature $T_N$, demonstrating a continuous suppression and elimination of AFM order under pressure.

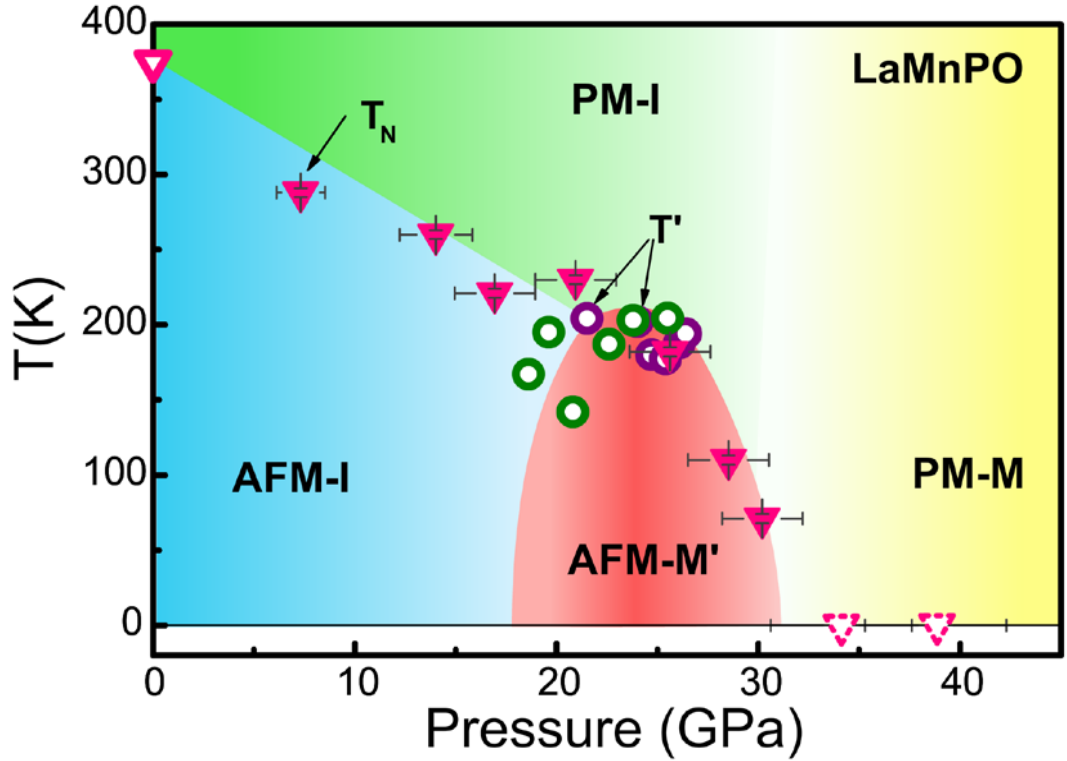

**Figure 4 The *P-T* electronic phase diagram of LaMnPO.** The open triangle represents the AFM transition temperature $T_N$ =375 K at ambient pressure, taken from neutron diffraction measurements[15,16]. The filled triangles correspond to values of $T_N$ obtained from our high pressure *ac* susceptibility measurements. The open circles represent temperatures T', where the resistance is a maximum, taken from different independent runs. There is no indication of AFM order above 32 GPa, as indicated by the dashed open triangles where $T_N$=0. The acronyms AFM-I and AFM-M' stand for antiferromagnetic insulating and antiferromagnetic mixed state regimes, respectively. PM-I and PM-M represent paramagnetic insulating and paramagnetic metallic states, respectively.